\documentclass[aps,prd,twocolumn,a4paper,showpacs,superscriptaddress,nofootinbib,floatfix]{revtex4}
\usepackage{latexsym}
\usepackage{graphics}
\usepackage{graphicx}
\usepackage{bm}
\usepackage{amsmath}
\usepackage{amsfonts}
\usepackage{amssymb}

\begin{document}
\newcommand{\missE}{E \hspace{-0.65em}/}
\newcommand{\pslash}{p\hspace{-2mm}/ \hspace{0mm}}
\newcommand{\be}{\begin{eqnarray}}
\newcommand{\ee}{\end{eqnarray}}
\newcommand{\nn}{\nonumber}
\newcommand{\ovl}{\overline}
\newcommand{\ra}{\rightarrow}
\newcommand{\lra}{\longleftrightarrow}
\newcommand{\ba}[1]{\begin{eqnarray} \label{(#1)}}
\newcommand{\ea}{\end{eqnarray}}
\newcommand{\rf}[1]{(\ref{(#1)})}
\newcommand{\dps}{\displaystyle}
\def\mNl{m_{\tilde{\nu}_\ell^N}}
\def\mM{\tilde{m}_M}
\def\0n{0\nu\beta\beta}
\def\LM{$L\hskip-2mm /\mbox{MSSM}\;$}
\def\mN{m_{\tilde{\nu}_\ell^N}}
\def\mD{\tilde{m}_D}
\def\Lbar{L\hskip-1.5mm/}
\def \lsim {\mbox{${}^< \hspace*{-7pt} _\sim$}}
\def \gsim {\mbox{${}^> \hspace*{-7pt} _\sim$}}
\def \leql { ^< \hspace*{-7pt} _=}
\def \geql { ^> \hspace*{-7pt} _=}
\def\rp{$R_p \hspace{-1em}/\;\:$}
\def\rpm{R_p \hspace{-0.8em}/\;\:}
\def\rpt{$R_p \hspace{-0.85em}/\ \ $}
\def\et{E_T \hspace{-1em}/}
\def\d{\partial \hspace{-0.55em}/}
\def\pmb#1{\setbox0=\hbox{#1}%
\kern-.015em\copy0\kern-\wd0
\kern.03em\copy0\kern-\wd0
\kern-.015em\raise.0233em\box0 }
\def\olrap{\overleftrightarrow{\partial}}
\def \znbb {0\nu\beta\beta}
\def \tnbb {2\nu\beta\beta}
\def \emass {\langle m_{\nu} \rangle}
\def\bfr{\pmb{${r}$}}
\def\bfsgm{\pmb{${\sigma}$}}
\def\sir{({ \bfsgm_{i}^{~}} \cdot {\hat{\bfr}_{ij}^{~}} )}
\def\sjr{({ \bfsgm_{j}^{~}} \cdot {\hat{\bfr}_{ij}^{~}} )}
\def\si{{ \bfsgm_{i}^{~}}}
\def\sj{{ \bfsgm_{j}^{~}} }
\hyphenation{re-la-ti-vis-tic}
\hyphenation{struc-ture}
\hyphenation{char-gi-nos}
\hyphenation{vio-la-ting}
\hyphenation{to-po-lo-gy}
\hyphenation{ava-i-lable}

\title{Lepton flavour violation in $\bm{e^{\pm}e^{-}\to  \ell^{\pm}e^{-}}$ 
($\bm{\ell = \mu,\tau}$) induced by R-conserving supersymmetry}   

\author{M.~Cannoni} 
\email[E-mail: ]{mirco.cannoni@pg.infn.it}
\affiliation{Dipartimento di Fisica, Universit\`a degli Studi di Perugia, Via A. Pascoli, I-06123, Perugia, Italy}
\affiliation{
Istituto Nazionale di Fisica Nucleare, Sezione di Perugia, Via A.~Pascoli, I-06123, Perugia, Italy}
\author{St.~Kolb}
\affiliation{
Istituto Nazionale di Fisica Nucleare, Sezione di Perugia, Via A.~Pascoli, I-06123, Perugia, Italy}
\author{O.~Panella}
\affiliation{
Istituto Nazionale di Fisica Nucleare, Sezione di Perugia, Via A.~Pascoli, I-06123, Perugia, Italy}

\date{18 June 2003}
\begin{abstract}{
The lepton flavour violating signals 
$e^{+}e^{-}\to  \ell^{+}e^{-}$ and 
$e^{-}e^{-}\to  \ell^{-}e^- \;(\ell=\mu,\tau)$ are 
studied in the context of low energy R-parity conserving supersymmetry   
at center of mass energies of interest for the next generation of 
linear colliders. Loop level amplitudes receive contributions from 
electroweak penguin and box diagrams involving sleptons and gauginos. 
Lepton flavour violation is due to off diagonal elements in 
$SU(2)_L$ doublet slepton mass
matrix. These masses are treated as model independent free phenomenological 
parameters in order to discover regions in parameter space where 
the signal cross
section may be observable. The results are compared with
(a) the experimental bounds from the non-observation of 
rare radiative lepton decays $\mu, \tau \to e\gamma$ and 
(b) the general mSUGRA theoretical scenario with seesaw mechanism 
where off diagonal slepton matrix entries are generated by 
renormalization group evolution of neutrino Yukawa 
couplings induced by the presence of new energy scales set by the heavy 
$SU(2)_L$ singlet neutrino masses. 
It is found that in $e^- e^-$ collisions the ($e\tau$) signal 
can be observable with a total integrated luminosity of 100 fb$^{-1}$ and 
the that the background can be easily suppressed. 
In  $e^+ e^-$ collisions the cross section is smaller and higher luminosities are needed. 
The experimental bound  on the decay $\mu \to e \gamma$ 
prevents the ($e\mu$) signal from being observable.
}
\end{abstract}


\maketitle

\section{Introduction}
\label{intro}

In the advent of growing evidence for neutrino oscillations and hence 
flavour mixing in the lepton sector of the Standard Model (SM), 
the topic of lepton flavour violation (LFV) has received considerable attention. 
Non-vanishing neutrino masses in principle induce LFV processes such as 
$\ell \rightarrow \ell' \gamma$. If neutrinos have masses in the 
eV or sub-eV range, the neutrino generated branching ratio to the latter
process is of order ${\cal O}$(10$^{-40}$) and therefore unobservably small.
On the other hand, in supersymmetric (SUSY) extensions of the SM the soft 
SUSY breaking potential $V_{soft}$ contains, in general, nondiagonal entries 
in generation space and therefore additional potential sources for LFV. 
Even in minimal supergravity scenarios characterized by universal 
soft mass term for scalar slepton and squark fields, renormalization 
induces potentially sizable weak scale flavour 
mixing~\cite{Gabbiani} in $V_{soft}$.

Many experimental efforts have been devoted to 
search for LFV and lepton number violating reactions, both 
in rare decays and in high energy accelerators. The strongest bounds on LFV 
come from the non-observation of radiative lepton 
decays~\cite{mega,cleo1,cleo2}:
\begin{eqnarray}
&&Br(\mu\to e\gamma)<1.2\times 10^{-11},\cr
&&Br(\tau\to e\gamma)<2.7\times 10^{-6},\cr 
&&Br(\tau\to \mu\gamma)<1.1\times 10^{-6}.
\label{boundexp}
\end{eqnarray}

The four LEP experiments searched for $Z\to \ell^{+}_{i}\ell^{-}_{j},\;
\ell=e,\mu,\tau, \;i\neq j$ at the Z peak providing 
the following upper bounds on branching ratios~\cite{pdg}: 
$Br(Z\to e\mu)<1.7\times 10^{-6}$,
$Br(Z\to e\tau)<9.8\times 10^{-6}$, 
$Br(Z\to \tau\mu)<1.2\times 10^{-5}$.
The high luminosity GigaZ option of the Tesla project~\cite{desy} 
is expected to probe the above branching ratios  
down to $\sim {\cal O}(10^{-8},10^{-9})$.
A recent study of LFV induced by R-parity conserving SUSY at the Z peak 
is given in Ref.~\cite{illa}.

However it is interesting 
to know if such signals can be observed at higher energies. 
The OPAL collaboration searched for LFV reactions 
up to the highest center of mass energy reached by LEPII, 
$\sqrt{s}= 209$ GeV~\cite{opal}. One  $e^{+}e^{-}\to e\mu$ event
was found at $\sqrt{s}=189$ GeV matching all tagging conditions, 
but it was interpreted as due to initial state radiation~\cite{opal}. 
This negative result implies the following upper limits 
(at $95\%$ confidence level) on the cross sections of LFV processes 
(for $200\; \mbox{GeV}\leq \sqrt{s} \leq 209\;\mbox{GeV}$): 
\begin{eqnarray}
&&\sigma(e^{+}e^{-}\to e\mu)<22\;\mbox{fb},\cr 
&&\sigma(e^{+}e^{-}\to e\tau)<78\;\mbox{fb},\cr
&&\sigma(e^{+}e^{-}\to \mu\tau)<64\;\mbox{fb}.
\label{sigmabounds}
\end{eqnarray}
For limits corresponding to lower energies see Ref.~\cite{opal}.

In the following this approach will be pursued further and a detailed study of 
the reactions
\begin{eqnarray} \label{reactions}
e^{+}e^{-}&\to&\ell^{+}e^- \ , \nonumber \\
e^{-}e^{-}&\to&\ell^{-}e^{-} \ (\ell=\mu,\tau)
\end{eqnarray} 
will be presented in the context of SUSY extension of the SM with conserved 
R-parity for center of mass energies of interest for the next linear 
collider projects (LC). The processes in Eq.~(\ref{reactions}) have the advantage of 
providing a clean final state being easy to identify experimentally (two back to 
back different flavor leptons), though one has to pay the price of dealing with 
cross sections of order $ {\cal O}(\alpha^4)$. Previous studies on SUSY
induced LFV at a LC (see {\it e.g.}~\cite{lfvee}) were limited to tree level processes 
for SUSY partner production decaying into final states characterized by very complicated 
topologies such as $\ell_{i} \ell_{j} +4 \,jets +\missE$ involving jets and missing energy. 
A detailed study of cuts and background is necessary to isolate the signal.

The relevant Feynmann diagrams describing the processes in Eq.~(\ref{reactions})
are shown in Figs.~\ref{boxep},~\ref{penguinep},~\ref{figee}. 
They  are the high energy analogue of the box and  
penguin diagrams that mediate LFV rare decays as e.g. $\mu \to e+\gamma$ or
$\mu \to 3e$. 
Due to the experimental limits on the cross sections and the loop nature of the 
process event rates are expected - even in more optimistic cases - to be relatively 
small. However, when the energy dependence of four-point and three-point 
functions is taken into account the amplitudes can show a 
resonance behavior as the energy approaches thresholds for particle 
production. This is a consequence of the discontinuity of the derivative 
of the real part of a loop amplitude where it develops an imaginary part 
(Cutkosky rule). The cross section in this point may increase by orders of magnitude. 
We have shown in a recent paper~\cite{Cannoni} on LFV induced by heavy Majorana neutrinos 
that the enhancement may be quite dramatic in some regions of the parameter space.

The plan of the paper is the following.
Sec.~\ref{sec2} discusses LFV in R-parity conserving 
SUSY and gives an outline of the calculation.  
Sec.~\ref{sec3} contains numerical results for the signal cross section 
and a discussion of possible backgrounds. 
Sec.~\ref{sec4} is devoted to a comparison with bounds from rare LFV 
lepton decays. Sec.~\ref{sec5} contains the conclusions. 
Appendices~\ref{lagrangian} and~\ref{appb} give details of the lagrangians 
and numerical tools used in the calculation.
Finally, in Appendix~\ref{amplitudes} helicity amplitudes for 
$e^{+}e^{-}$ and $e^{-}e^{-}$ collisions are given.     

\section{SUSY origin of lepton flavour violation}
\label{sec2}
One of the most important challenges in contemporary particle physics 
is to understand the origin of neutrino masses. Quite generally this 
requires new fields to be added to those of the SM and/or those of its 
minimal SUSY version (MSSM). In the seesaw framework - the simplest scenario for the  
explanation of neutrino masses - and its SUSY extension, the superpotential
contains three $SU(2)_L$ singlet neutrino superfields $N_{i}$ with the following 
couplings~\cite{borma,hisa1,hisa2,casas,hisa3}:
\begin{eqnarray}
W=(Y_{\nu})_{ij}\varepsilon_{\alpha\beta}H_{2}^{\alpha}N_i L^{\beta}_j
+\frac{1}{2}(M_{R})_i N_i N_i.
\label{yuk}
\end{eqnarray}
Here $H_2$ is a Higgs doublet superfield, $L_i$ are the $SU(2)_L$ doublet lepton superfields, 
$Y_{\nu}$ is a Yukawa coupling matrix and $M_R$ is the $SU(2)_L$ singlet neutrino mass matrix.
As is usually done the basis has been chosen such that $M_R$ is diagonal. The effective low 
energy neutrino mass matrix is given by
\begin{eqnarray}
{\cal M}_{\nu}=m_{D}^{T}M_{R}^{-1}m_{D},
\end{eqnarray}
where $m_{D}=v_{2}(Y_{\nu})_{ij}/{\sqrt 2}$ is the Dirac neutrino mass matrix  
and $v_{2}=\langle H_{2}^{0}\rangle$. 

Standard mSUGRA models contain an universal
GUT scale ({\it i.e.} at the energy scale where the coupling constants unify)
scalar field mass term $m_{0}$. At low energies 
the renormalization group equations (RGE) produce diagonal slepton mass matrices. 
With the additional Yukawa couplings in Eq.~(\ref{yuk}) and a new mass scale ($M_R$) 
the RGE evolution of the parameters is modified: assuming that $M_{R}$ is the mass scale 
of heavy right-handed neutrinos, the RGE from GUT scale to $M_R$ induce
off-diagonal matrix elements in  $(m^{2}_{\tilde{L}})_{ij}$. In the one 
loop approximation the off-diagonal elements are~\cite{casas}:
\begin{eqnarray}
(m^{2}_{\tilde{L}})_{ij}\simeq -\frac{1}{8\pi^{2}}(3+a^{2}_{0})m_{0}^{2}
(Y_{\nu}^{\dagger}Y_{\nu})_{ij}\ln\left(\frac{M_{GUT}}{M_{R}}\right).
\label{dl2}
\end{eqnarray} 
Here $a_{0}$ is a dimensionless parameter appearing in the matrix of 
trilinear mass terms 
$A_{\ell}=Y_{\ell}a_{0}m_{0}$ contained in $V_{soft}$.

The slepton mass eigenstates are obtained diagonalizing the slepton 
mass matrices. 
The corresponding mixing matrices induce LFV couplings in the 
lepton-slepton-gaugino vertices  
$\tilde{\ell}^{\dagger}_{L_{i}}{U_{L}}_{ij}\tilde{\ell}_{L_{j}}\chi$. 
The same effect on the
mass matrix of $SU(2)_L$ singlet charged 
sleptons $(m^{2}_{\tilde{R}})_{ij}$ is 
smaller~\cite{casas,hisa3}.

The magnitude of LFV effects will depend on the RGE induced non diagonal entries and 
ultimately on the neutrino Yukawa couplings $(Y_{\nu})_{ij}$. These in turn depend 
on the fundamental theory in which this mechanism 
is embedded (for example $SU(5)$ or $SO(10)$ SUSY GUT~\cite{hisa3,biqi,masiso10}) and on the 
particular choice of texture for the neutrino mass matrix~\cite{casas,ellis,pas}. 
The rate of LFV transitions like $\ell_{i} \to \ell_{j}$, $i\neq j$, 
$\ell=e,\mu,\tau$ induced by the lepton-slepton-gaugino vertex is determined by the
mixing matrix ${U_{L}}_{ij}$ that, as stated above, is model dependent.
In a model independent way, however, one can take the 
lepton, slepton, gaugino vertex flavour conserving with the slepton 
in gauge eigenstates, so that LFV is 
given by mass insertion of non diagonal slepton propagators~\cite{Gabbiani,illa,hisa1}. 

In a similar spirit, the phenomenological study presented in this paper
will be quite model independent and in order to keep the discussion simple 
the mixing of only two generations is considered, so that the slepton and sneutrino 
mass matrix is:

\begin{eqnarray}
\tilde{m}^{2}_{{L}}=\left(\begin{array}{cc}
 \tilde{m}^{2} & \Delta m^2\\
       \Delta m^2 & \tilde{m}^{2}
\end{array}\right),
\end{eqnarray}
with eigenvalues: $\tilde{m}^{2}_{\pm}=\tilde{m}^2\pm \Delta m^2$ 
and maximal mixing matrix  
\begin{eqnarray}
U=\frac{1}{\sqrt{2}}
\left(\begin{array}{cc} 
     1 & 1 \\
     1 & -1
\end{array}\right).
\end{eqnarray}
Under these assumptions the LFV propagator in momentum space for a scalar line is
\begin{eqnarray}
\langle\tilde{\ell}_{i}\tilde{\ell}^{\dagger}_{j}\rangle_{0}&=&
\frac{i}{2}\left(
\frac{1}{p^{2}-\tilde{m}^{2}_{+}}-\frac{1}{p^{2}-\tilde{m}^{2}_{-}}\right)\cr
&=&i
\frac{\Delta m^2}{(p^{2}-\tilde{m}^{2}_{+})(p^{2}-\tilde{m}^{2}_{-})} \ ,
\label{LFVprop} 
\end{eqnarray} 
while a lepton flavour conserving (LFC) scalar line is described by
\begin{eqnarray}
\langle\tilde{\ell}_{i}\tilde{\ell}^{\dagger}_{i}\rangle_{0}=
\frac{i}{2}\left(
\frac{1}{p^{2}-\tilde{m}^{2}_{+}}+\frac{1}{p^{2}-\tilde{m}^{2}_{-}}\right).
\label{LFCprop} 
\end{eqnarray} 
Therefore the essential parameter that controls the LFV signal is
\be 
\delta_{LL}=\frac{\Delta m^{2}}{\tilde{m}^{2}}.
\label{lfvpar}
\ee
Before presenting detailed calculations a qualititave order 
of magnitude estimate of the cross section can be given using dimensional 
arguments. Consider for simplicity a box diagram. Neglecting the external momenta 
in the loop propagators and indicating with $m_S$ a typical SUSY mass, one has for the amplitude in the case of a scalar four
point function:
\be
{\cal M}\simeq \frac{g^4}{(4\pi)^2}\,s\, m_{S}^{2}\, 
\frac{m_{S}^{2}{\Delta m^2}}{m_{S}^{8}}.
\ee
The constant comes from couplings and loop integration, the factor $s$
from the spinorial part, the mass squared factor from the numerator of the 
two gaugino propagators and the last factor from the loop integral.
The corresponding total cross section 
(assuming polarized initial particles) is:
\be
\sigma \simeq \frac{1}{16\pi}\left(\frac{\alpha}{\sin^{2}\theta_{W}}\right)^{4}
\delta^{2}_{LL}\frac{s}{m^4_S}.
\ee  
Taking $m_{S}=100$ GeV, $\delta_{LL}=0.1$ and $\sqrt{s}=200$ GeV one has
$\sigma \simeq 1.3\times 10^{-2}$ fb while with $\sqrt{s}=500$ GeV
$\sigma \simeq 8$ fb. With an annual integrated luminosity of order $L_0$=100 fb$^{-1}$
one may expect an observable signal.

However this estimate is clearly too crude: it gives a linear increase with 
$s$ while one expects at high energies, $\sqrt{s}>> m_S $, a cross section which scales 
as $s^{-1}$. To get a realistic result it is necessary to compute exactly the energy 
dependence of the loop integrals and the interference among all 
contributing graphs.

\section{Numerical results}
\label{sec3}

\begin{figure}
\begin{center}
\scalebox{0.45}{\includegraphics*[75,130][510,780]{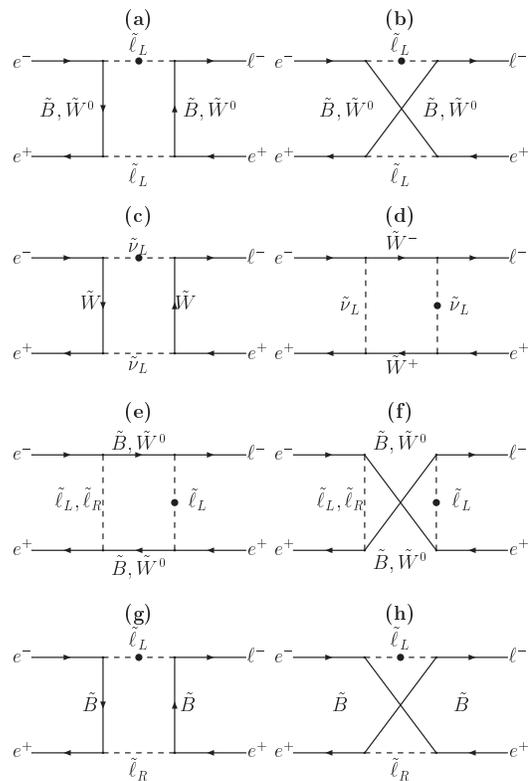}}
\caption{Box diagrams for $e^{+}e^{-}$ collisions. The full black dot 
in a scalar line denotes the lepton flavour violating 
propagator (Eq.~\protect\ref{LFVprop}). 
}
\label{boxep}
\end{center}       
\end{figure}
\begin{figure}
\begin{center}
\scalebox{0.45}{\includegraphics*[50,110][550,780]{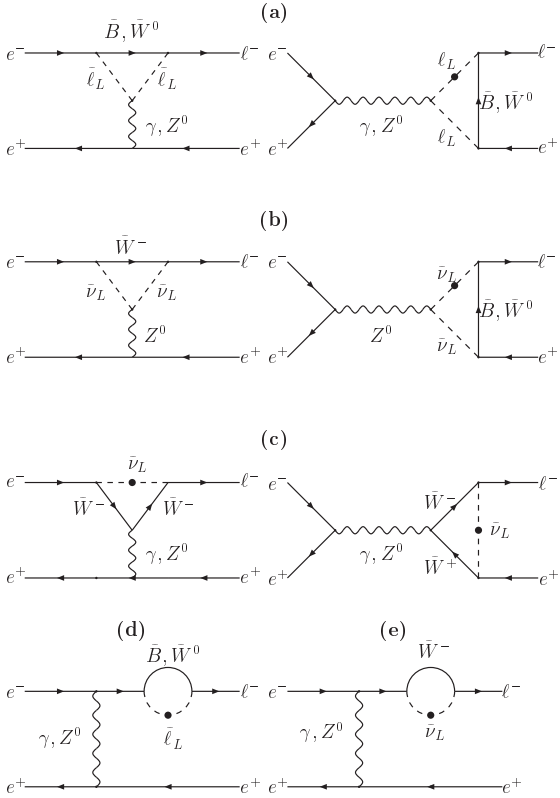}}
\caption{Penguin and external legs diagrams for $e^{+}e^{-}$ collisions. 
The full black dot 
in a scalar line denotes the lepton flavour violating propagator. 
In the diagrams where it is not marked it can occour in both lines.
Diagrams like (d) and (e) with the gauge boson in the 
$s$ channel are  also present but not shown.}
\label{penguinep}
\end{center}       
\end{figure}
\begin{figure}
\begin{center}
\scalebox{0.45}{\includegraphics*[50,110][550,780]{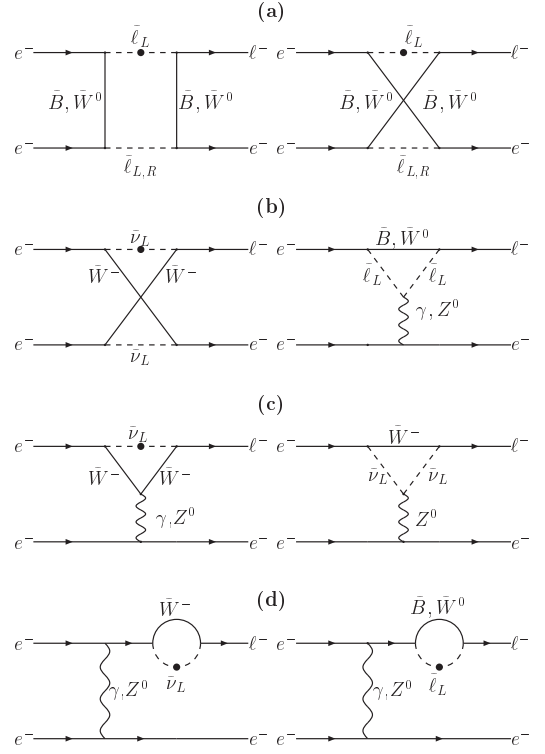}}
\caption{Feynman diagrams for $e^{-}e^{-}$ collisions. The full black dot 
in a scalar line denotes again the lepton flavour violating propagator. 
Exchange diagrams are not shown.}
\label{figee}
\end{center}       
\end{figure}
In the reactions considered here there are only leptons in the initial and
final state. At the energies of a LC lepton masses 
can be safely neglected and thus all the calculations are 
done assuming massless external fermions. 
The signal is suppressed if neutralinos and charginos $\chi^{0,\pm}$ 
are Higgsino-like, since their coupling is 
proportional to the lepton masses. 
For the same reason left-right mixing in the slepton matrix is neglected.
Therefore it is assumed that
the two lightest neutralinos are pure Bino and pure Wino 
with masses $M_1$ and $M_2$ respectively, while 
charginos are pure charged Winos with mass $M_2$, $M_1$ and $M_2$  
being the gaugino masses in the soft breaking potential.
The relevant parts of the interaction lagrangian are listed in 
Appendix~\ref{lagrangian}.  

Due to the chiral nature of the couplings it is convenient
to calculate the amplitudes using the helicity base for spinors: 
the amplitudes are written in terms of spinor pro\-ducts and a 
numerical code can be easly implemented to compute both real 
and imaginary parts. Interference terms are 
also accounted for by summing the various contributions before 
taking the absolute modulus squared of the amplitude.
In the helicity  basis and in the limit of massless fermions 
there are only two independent spinors, 
$u_{+}(p)\equiv u_{R}(p)$ and $u_{-}(p)\equiv u_{L}(p)$ with 
only two non-zero spinor products: 
$\ovl{u}_R(p_a) u_L(p_b)\equiv S(p_a,p_b)$, 
$\ovl{u}_L(p_a) u_R(p_b)\equiv T(p_a,p_b)$ given by compact 
expressions, see Appendix~\ref{appb1}. 
The loop integrals are decomposed in form factors and calculated 
numerically using the package {\scshape{LoopTools}} ~\cite{looptools}. 
The decomposition of loop integrals is obtained for 
massless external particles and with the loop momenta assigned as described
in Fig.~\ref{loopkin} of Appendix~\ref{appb2}. The exact 
dependence from the masses of the particles exchanged in the loop 
is also given in Appendix~\ref{appb2}. Assigning the momenta in a different 
way corresponds to a shift of the integration variables and produces different 
combinations of the loop form factors appearing in the amplitudes. 
The numerical 
values remain unchanged.

Besides computational advantages the helicity method 
clarifies the physics of the processes.
The momenta of the external particles are specified as in 
Eq.~(\ref{momenta}) (Appendix B) and Fig.~\ref{loopkin} (Appendix C), 
and the following reactions are considered:
\be
e^+(p_1,\lambda_1)e^{-}(p_{2},\lambda_2)&\to& 
\ell^{-}(p_3,\lambda_3)e^+(p_4,\lambda_4),\cr 
e^-(p_1,\lambda_1)e^{-}(p_{2},\lambda_2)&\to& 
\ell^{-}(p_3,\lambda_3)e^-(p_4,\lambda_4).
\ee 
Here $\lambda_i$ denotes the helicity of particle $i$. 
The corresponding helicity amplitudes ${\cal M}_j$ expressed in terms 
of spinor pro\-ducts
and {\scshape{LoopTools}} form factors are obtained after tedious but 
straightforward algebra. They can be found in Appendix~\ref{amplitudes}. 

The integrated cross sections corresponding to each individual amplitude 
${\cal M}_j$ is:
\be
\sigma_j&=&\frac{1}{32\pi s}\int d(\cos{\theta})\,
|{\cal M}_{j}|^{2}. 
\ee
The total unpolarized cross-section (averaged over initial spins) is $\sigma=(1/4)\sum_j \sigma_j$.
The dependence on the scattering angle is encoded in the Mandelstam 
variables $u$ and $t$. Numerical results are obtained using the mSUGRA relation 
$M_1 \simeq 0.5 M_2$ for gaugino masses while $\Delta m^{2}$ and the slepton 
masses are taken to be free phenomenological parameters. 
The parameter space is scanned in order to identify the regions which may 
deliver an interesting signal. The discussion of whether such regions are 
compatible with present experimental bounds is postponed to the next section.

\subsection{$\bm{e^+e^-}$ collisions}

\begin{figure}
\begin{center}
\scalebox{0.45}{\includegraphics*[0,150][600,700]{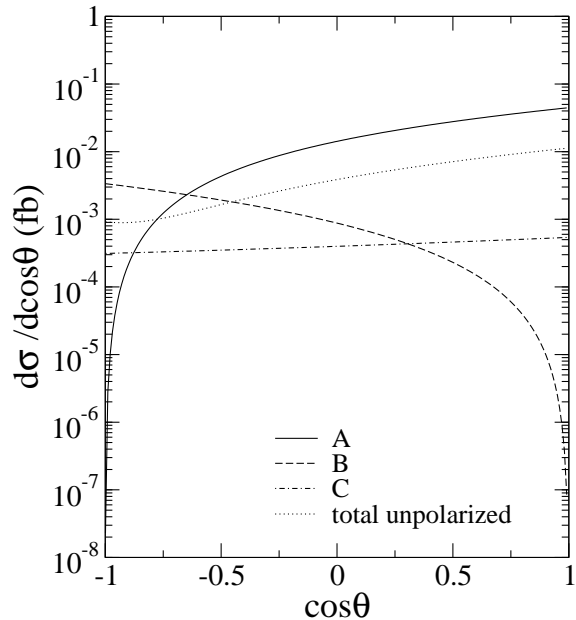}}
\caption{Differential cross section as a function of the scattering angle
for $e^{+}e^{-}$ collisions. The following values of the parameters are used:
$M_1=80,\; M_2=160,\;m_{\tilde{\ell}}=m_{\tilde{\nu}}=100$ GeV and 
$\Delta m^{2}=6000$ $\mbox{ GeV}^2$.}   
\label{pediff}
\end{center}       
\end{figure}
\begin{figure}
\begin{center}
\scalebox{0.5}{\includegraphics*[50,90][580,800]{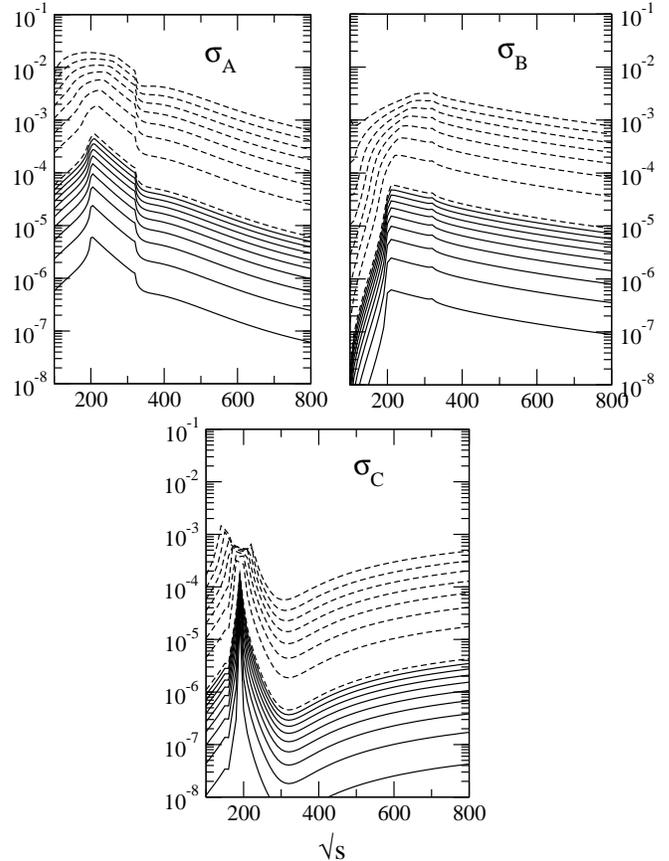}}
\caption{Total cross section (fb) as a function of $\sqrt{s}$
for $e^{+}e^{-}$ collisions for the three helicity amplitude. The 
parameters chosen are $M_1=80,\ M_2=160,\ m_{\tilde{\ell}}=m_{\tilde{\nu}}=100$ GeV.
The solid lines correspond to $\Delta m^2$ increasing from $100$ GeV$^{2}$
to $900$ GeV$^{2}$ in steps of $100$. The dashed lines correspond to 
$\Delta m^2$ increasing from $1000$ to $8000$ GeV$^{2}$ in steps of $1000$.}   
\label{petot}
\end{center}       
\end{figure}
The contributing amplitudes are (Appendix~\ref{suba}):
\be
{\cal M}_A&=&{\cal M}(e^+_R e^-_L \to \ell^-_L e^+_R),\cr
{\cal M}_B&=&{\cal M}(e^+_L e^-_R \to \ell^-_L e^+_R),\cr
{\cal M}_C&=&{\cal M}(e^+_L e^-_L \to \ell^-_L e^+_L).
\ee
For each helicity amplitude the corresponding differential polarized
cross section is shown in Fig.~\ref{pediff}. 
The different behavior is easily understood in terms of 
helicity conservation at high energy.
{\em Amplitude} ${\cal M}_A$ is peaked in the forward direction since   
it has a P-wave initial state with $J_z=+1$. Angular 
momentum conservation requires the right-handed positron to be emitted 
in the positive direction of the collision axis while the left-handed 
negative charged lepton must have its momentum in the opposite direction.   
{\em Amplitude} ${\cal M}_B$ is peaked in the backward direction as 
it is a P-wave scattering with $J_z=-1$. The right-handed positron must be 
emitted backward while the negative charged lepton is in the forward 
direction. 
{\em Amplitude} ${\cal M}_C$ has no virtual vector boson exchanged and is 
an S-wave ($J_z=0$) scattering. One expects therefore an almost flat, isotropic 
distribution.

The dominating contribution to the integrated unpolarized cross section
comes from amplitude 
${\cal M}_A$, which is an order of magnitude larger than ${\cal M}_B$ and 
two orders of magnitude larger than ${\cal M}_C$ in most of the phase space. 
Only for large scattering angles (backward direction) 
the amplitude ${\cal M}_B$ 
dominates and ${\cal M}_A$ is the smallest one. In Fig.~\ref{pediff} the 
dotted line corresponds to the unpolarized differential cross section 
({\it i.e.} the inchoerent sum of the contributions of ${\cal M}_{A,B,C}$ 
averaged over the initial spins). 
It is worth to remark that in such circumstances the possibility 
of having polarized electron and positron beams would maximize the chances
to observe these signals. Considering the unpolarized cross section
corresponds essentially to calculating 
$\sigma_{unpol} \approx (1/4)\, \sigma(e^+_R e^-_L \to \ell^-_L e^+_R)$.

Fig.~\ref{petot} shows the cross section integrated over the scattering angle 
for the three helicity amplitude 
as a function of the  center of mass energy $\sqrt{s}$ and for 
increasing values of the LFV parameter 
$\Delta m^2$. The presence of spikes
is due to the onset of the absorptive part of the diagrams  
corresponding to thresholds of real particle pair production. 
For the values of masses used in Fig.~\ref{petot} one expects 
thresholds effects at $\sim 200$ GeV for slepton pair production 
and $\sim 320$ GeV for gaugino pair production. This is evident for in $\sigma_A$ (upper-left panel)
and $\sigma_B$ (upper-right panel). The shape is determined in the first case by the 
destructive interference among the two types of box graphs (with scalars and fermions on threshold) 
and by the value of 
$\Delta m^2$ inducing two distinct thresholds at $\tilde{m}^{2} \pm \Delta m^2$. 
$\sigma_B$ is determined only by penguin diagrams that give smaller contribution 
relative to the boxes. 
$\sigma_C$ receives contributions only from box diagrams: at threshold for sleptons production 
its value varies by orders of magnitude differently from the two other cases. 
This can be easily understood considering the threshold behavior of the 
cross section for sleptons pair production~\cite{Peskin}: 
defining $\beta$ the selectron velocity, 
the intermediate state of the amplitudes ${\cal M}_A$ and ${\cal M}_B$ 
correspond to the 
reactions $e^-_L e^+_R \to \tilde{e}^-_L \tilde{e}^+_L$ 
and $e^-_R e^+_L \to \tilde{e}^-_L \tilde{e}^+_L$
that near threshold behave like $\beta^3$ while amplitude ${\cal M}_C$ 
corresponds to the reaction $e^-_L e^+_L \to \tilde{e}^-_L \tilde{e}^+_R$ 
that at threshold behaves like $\beta$. 

\begin{figure}
\begin{center}
\scalebox{0.45}{\includegraphics*[20,150][600,800]{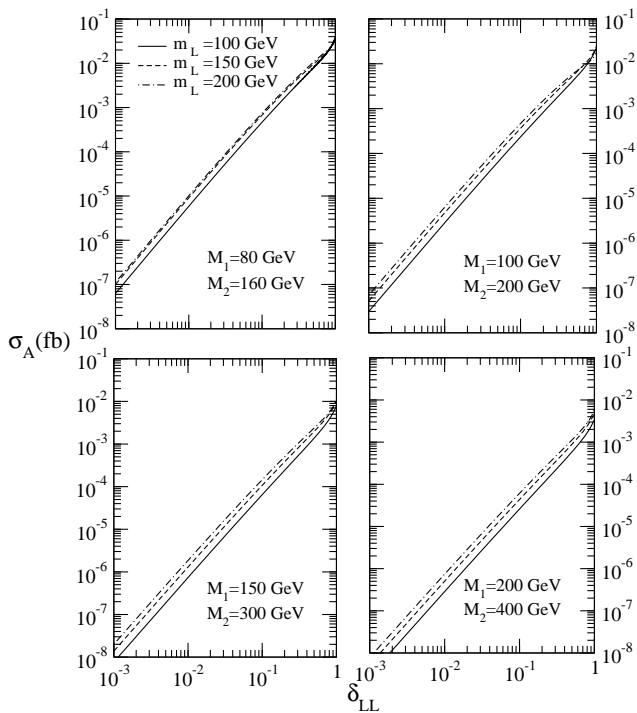}}
\caption{Total cross section for the amplitude A as a function of 
the dimensionless parameter $\delta_{LL}$ (see Eq.~(\ref{lfvpar})) and for
$\sqrt{s}=2\tilde{m}_L$. The values of the other parameters are given in the 
legends.} 
\label{pedelta}
\end{center}       
\end{figure}

The cross section is peaked around $\sqrt{s}=2 \tilde{m}_L$. 
In Fig.~\ref{pedelta} $\sigma_A=\sigma(e_R^+e_L^- \to \ell^-_L e^+_R)$ 
is shown as a function of 
$\delta_{LL}$ for $\sqrt{s}=2\tilde{m}_L$ and for different values of slepton 
and gaugino masses. Given an annual integrated luminosity $L_0$=100 $fb^{-1}$ 
a cross section of $10^{-2}$fb produces one signal event per year. Such an event rate 
is reached only for $M_1$ not larger than $\sim 200$ GeV and 
$\delta_{LL}\simeq {\cal O}(1)$. This hypothesis will be discussed in the next section.     
Moreover angular cuts in the forward direction are needed to suppress possible SM
backgrounds and - since the largest values of the cross section correspond to small
angles - the signal will be affected by such a cut. Therefore the observation
of LFV in $e^+ e^-$ collisons appears to be difficult unless $L_0$ 
is considerably larger than $10^2$ fb$^{-1}$/yr. 

\subsection{$\bm{e^-e^-}$ collisions}

\begin{figure}
\begin{center}
\scalebox{0.45}{\includegraphics*[0,150][600,700]{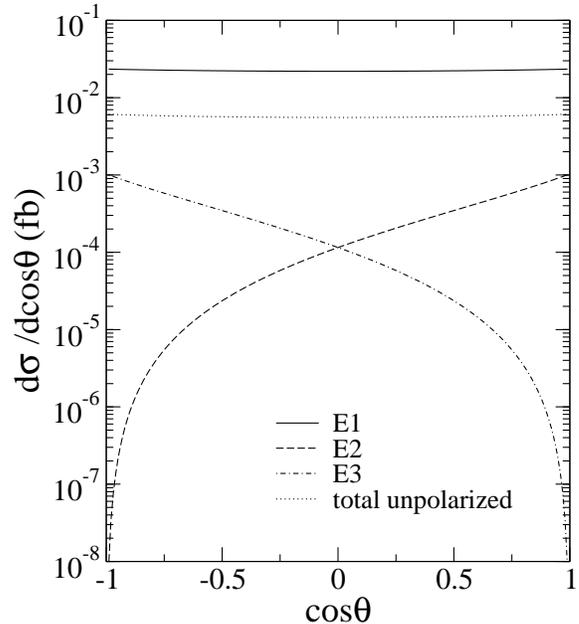}}
\caption{Differential cross section as a function of the scattering angle
for $e^{-}e^{-}$ collisions. The choice of the parameters is the same as in 
Fig.~\ref{pediff}.}
\label{eediff}
\end{center}       
\end{figure}
The contributing amplitudes are (Appendix~\ref{subb})
\be
{\cal M}_{E1}&=&{\cal M}(e^-_L e^-_L \to \ell^-_L e^-_L),\cr
{\cal M}_{E2}&=&{\cal M}(e^-_L e^-_R \to \ell^-_L e^-_R),\cr
{\cal M}_{E3}&=&{\cal M}(e^-_R e^-_L \to \ell^-_L e^-_R).
\ee
The corresponding differential cross sections are plotted in 
Fig.~\ref{eediff}. 
${\cal M}_{E1}$ has $J_z=0$, is flat and forward-backward 
symmetric because of the antisymmtrizzation.
${\cal M}_{E2}$ and ${\cal M}_{E3}$ describe
P-wave scattering with $J_z=+1$ and $J_z=-1$ respectively: in order to 
conserve angular momentum ${\cal M}_{E2}$ must be peaked in the 
forward direction while ${\cal M}_{E3}$ favours backward scattering.
Both ${\cal M}_{E2}$ and ${\cal M}_{E3}$ 
are orders of magnitude smaller than ${\cal M}_{E1}$. 
The signal cross section is to a very good approximation given 
by the amplitude ${\cal M}_{E1}$. Since it is almost flat
the angular integration will give a factor almost exactly equal to two. 
This again shows  the importance of the option of having polarized beams.
If both colliding electrons are left-handed one singles out 
the dominant helicity amplitude and a factor four is gained
in the cross section relative to the unpolarized case.
This may be important in view of the relatively small signal 
cross section one is dealing with.
\begin{figure}
\begin{center}
\scalebox{0.45}{\includegraphics*[0,150][600,700]{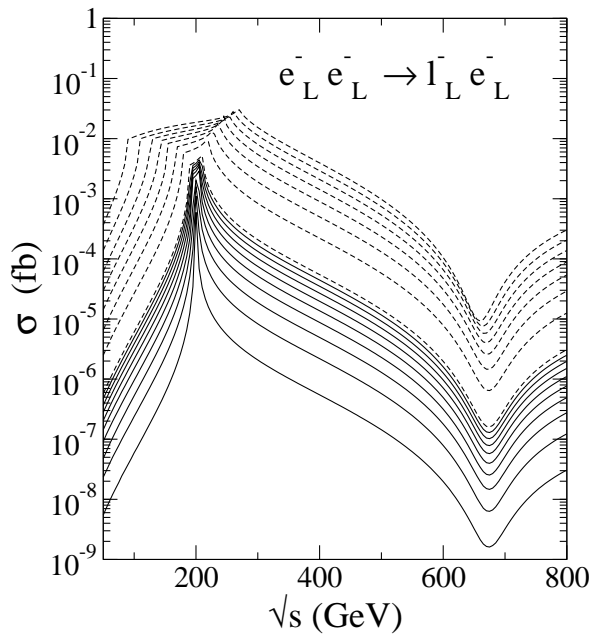}}
\caption{Total cross section of the dominant amplitude E1 as a function of $\sqrt{s}$
for $e^{-}e^{-}$ collisions. The choice of the parameters is the same as in 
Fig.~\ref{pediff}.}
\label{eetot}
\end{center}       
\end{figure}
In this case, due to the smaller number of diagrams, the analysis of the total 
cross section as a function of $\sqrt{s}$ is easier (see Fig.~\ref{eetot}): 
the box diagrams dominate
at $\sqrt{s}=2\tilde{m}_L$ where $\sigma$ changes of orders of magnitude 
giving a sharp peak that is smeared only by large values of 
$\Delta m^2$, while penguin diagrams give a substantial contribution only at 
higher energies.
The reason is the same as for the $\sigma_C$ behaviour in the $e^+ e^-$ case: 
the intermediate state $e^-_L e^-_L \to \tilde{e}^-_L \tilde{e}^-_L$ behaves like 
$\beta$, while the other two like $\beta^3$.
Here the highest absolute value is due to the couplings and the constructive 
interference of boxes where both Binos and Winos can be exchanged.
\begin{figure}
\begin{center}
\scalebox{0.45}{\includegraphics*[20,180][600,800]{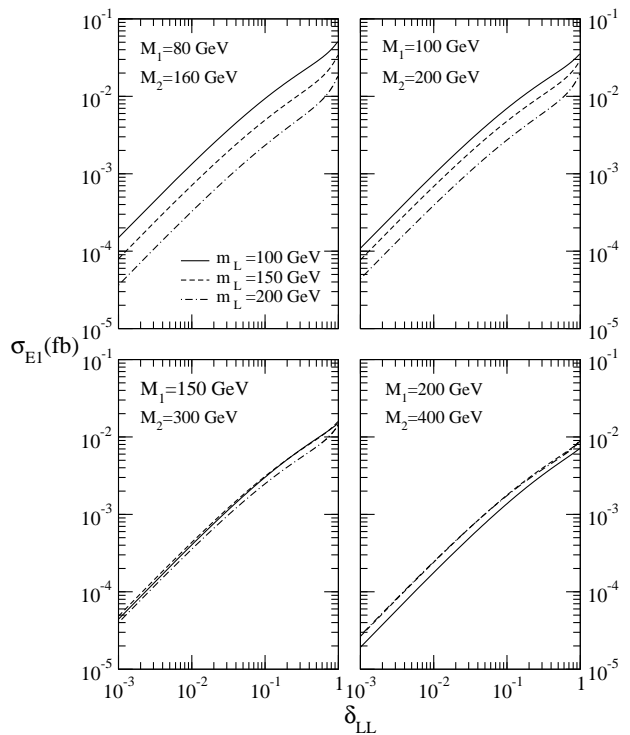}}
\caption{Total cross section for the amplitude E1 in function of 
the dimensionless parameter $\delta_{LL}$, see Eq.~(\ref{lfvpar}). The values 
of the other parameters are given in the legends. Each plotted line is 
calculated assuming $\sqrt{s}=2\tilde{m}_L$.}
\label{eedelta}
\end{center}       
\end{figure}
The dependence of $\sigma_{E1}$ on $\delta_{LL}$ is shown in Fig.~\ref{eedelta}. 
With SUSY masses not much larger than $\sim 200$ GeV the signal is of order 
${\cal O}$(10$^{-2}$) fb for $\delta_{LL} > {\cal O}(10^{-1})$. Relative to 
the $e^{+}e^{-}$ case there are 
two important features: $(i)$ the cross section is practically angle independent so 
that it is insensitive to angular (or tranverse momentum) cuts; $(ii)$ the SM 
background - though not completely absent - can be easily controlled as will be
shown in the next subsection.

\subsubsection{$\bm{Background}$}    
The signal has the unique characteristic of a back to back high 
energy lepton pair. Sources of background were qualitatively discussed in 
Ref.~\cite{Heusch}.

Initial and final state radiation can  
be a source of background. An example is the OPAL event~\cite{opal}, 
although lepton pairs can hardly have  
the same kinematical feature of the signal.  
Other sources present multiparticle final
states (at least six particles) and missing energy due to the presence of
neutrino pairs. 

The first type is given by reactions like $e^{-}e^{-} \to e^{-}e^{-} b^{*} {\bar{b}}^{*}$
that proceedes trough virtual photon fusion. The subsequent chain of weak deacys 
produces a final state with missing momentum, 
hadronic jets and opposite or same sign leptons, that however can be 
again separated using the clear kinematical topology of the signal. 

\begin{figure}
\begin{center}
\scalebox{0.45}{\includegraphics*[35,195][550,785]{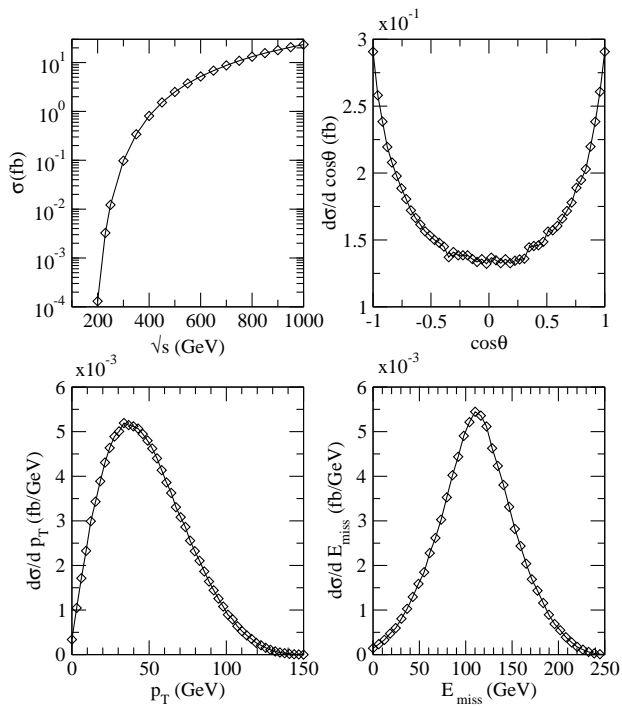}}
\caption{
Total cross section and distributions for 
$e^{-}e^{-}\to W^{-}W^{-}\nu \nu$ 
Upper-left figure: total cross section as a function of $\sqrt{s}$. 
Upper-right: angular distribution for 
a $W^{-}$ where $\theta$ is the angle among the collision axis and the boson momentum. 
Bottom-left: distribution of the transverse momentum of $W^{-}$.
Bottom-right: energy distribution of the two neutrinos. All distributions are calculated 
with $\sqrt{s}=300$ GeV.}
\label{backg}
\end{center}       
\end{figure}

A second type:
\be
&&e^{-}e^{-} \to \nu_{e}\nu_{e} {W^{-}}^{*} {{W}^{-}}^{*}\cr
&&{W^{-}}^{*} {{W}^{-}}^{*} \to 
\ell^{-} {\bar{\nu}}_{\ell}\,{\ell^{-}}' {\bar{\nu}}_{{\ell}'}
\ee
with four neutrinos and a like sign-dilepton pair that can be of the same or
different flavour. 
This appears to be the most dangerous background, as it produces two 
leptons and missing energy, and therefore it is analyzed in more detail.
Moreover, to the best of our knowledge, it has not been previously 
considered in the literature. Fig.~\ref{backg} shows the total 
cross section $e^{-}e^{-} \to \nu_{e}\nu_{e} {W^{-}}{{W}^{-}}$ calculated 
with the {\scshape{CompHEP}} package~\cite{comphep}, that allows to 
compute numerically the $17$ Feynman diagrams contributing at tree level.
Above the threshold for $W^{-}W^-$ gauge boson
production the cross section rises rapidly by orders of magnitude, 
becoming almost constant at high energies. 
In the region $\sqrt{s}\simeq 250-400$ GeV it increases from $10^{-2}$ 
fb to $1$ fb. In order to get an estimate of the cross section for   
the six particle final state process, the cross section 
$\sigma(e^{-}e^{-} \to {W^{-}}{{W}^{-}}\nu \nu)$ 
has to be multiplied by the branching ratio of the leptonic decays of 
the two gauge bosons, $\simeq 10\%$, so that  
$\sigma_{Background}\simeq 10^{-4}-10^{-2}$ fb, and it is at the level of the signal.
However the kinematical configuration of the final state 
leptons is completely different.    
Fig.~\ref{backg} (upper-right) shows the angular distribution
of the gauge bosons which is peaked in the forward and backward 
directions so that the leptons produced in the $W$ gauge boson decay 
are emitted preferentially along the collsion axis.
In addition their tansverse momenta will be softer compared to that of 
the signal: Fig.~\ref{backg} (bottom-left panel) shows that the 
transverse momenta distribution of the gauge bosons is peaked 
at $p^{P}_T=(\sqrt{s}/2-M_W)/2 \simeq 35$ GeV for 
$\sqrt{s}=300$ GeV. Consequentely the lepton distributions
will be peaked at $p^{P}_T/2 \simeq 17.5$ GeV.
The missing energy due to the undetected neutrinos
(Fig.~\ref{backg}, bottom-right panel) can be as large as 
$\simeq \sqrt{s} - 2M_W$. This distribution should be convoluted with
that of the neutrinos produced in the gauge boson decay.
Therefore it can be safely concluded that it will be possible to control 
this background because, with reasonable cuts on the transverse momenta 
and missing energy, it will be drastically reduced, while - as mentioned
above - these cuts will not affect significantly the signal.

\section{Comparison with rare lepton radiative decays}  
\label{sec4}

The main result of the calculations presented in the previous
sections is that, as it can be inferred from Fig.~\ref{eedelta},  
the phenomenological points of the SUSY parameter space corresponding to gaugino 
masses $(M_1,\;M_2)=(80,\;160)$ GeV or  $(100,\;200)$ GeV and to 
slepton masses $m_L=100-200$ GeV and $\delta_{LL}>10^{-1}$
(which implies $\Delta m^2>10^{3}$ GeV$^{2}$) can give in the 
$e^- e^-$ mode a detectable LFV signal ($e^-e^- \to \ell^- e^-$)  
although at the level of ${\cal O}(1-25)$ events/yr with 
${L}_0=100$ fb$^{-1}$. Higher sensitivity to the SUSY 
parameter space could be obtained with larger $L_0$. 
It is interesting to note that this light sparticles spectrum 
that is promising for collider discovery, is also preferred 
by the electroweak data fit. In Ref.~\cite{Altarelli} it is 
shown that light sneutrinos, charged left sleptons and light gauginos 
improve the agreement among the electroweak precision measurements 
and the lower bounds on the Higgs mass.

\begin{figure}
\begin{center}
\scalebox{0.55}{\includegraphics*[110,140][545,755]{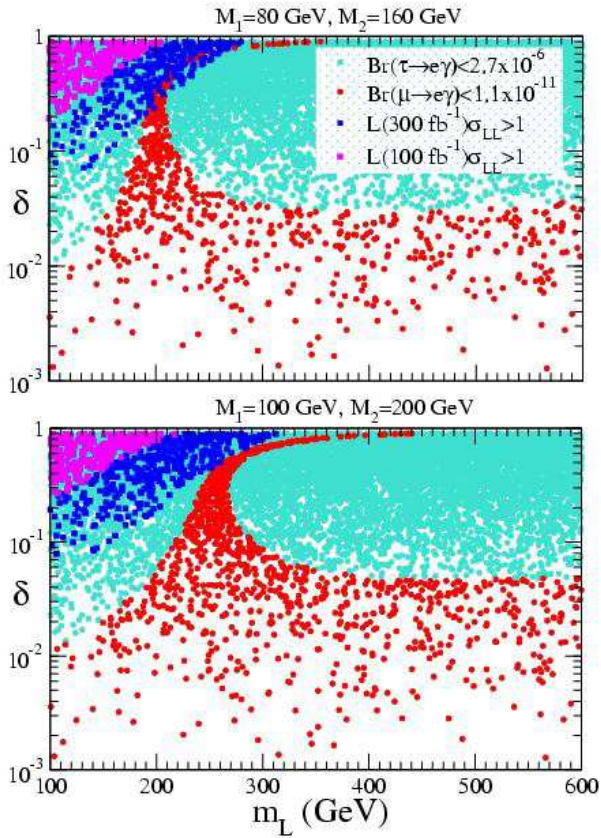}}
\caption{
Scatter plot in the plane ($\delta_{LL}, m_L$) of: (a) the 
experimental bounds from $\mu\to e\gamma$ and 
$\tau \to \mu \gamma$ (allowed regions with circular dots); (b) regions where the signal 
can give at least one event with two different values of integrated 
luminosity (squared dots), for two sets of gaugino masses. Each signal point is calculated 
at $\sqrt{s}=2\tilde{m}_L$.
}
\label{scaplot}
\end{center}
\end{figure}

On the other hand the experimental bounds on rare lepton 
decays set constraints on the LFV violating paramters $\Delta \tilde{m}^2$ or $\delta_{LL}$:
the constraints in Eq.~(\ref{boundexp}) define an allowed 
(and an excluded) region in the plane ($\delta_{LL}, m_L$) 
which are computed using the formulas given in Ref.~\cite{hisa2} (adapted to our model)  
for the LFV radiative lepton decays. These regions have to be compared 
with those satisfying the ``discovery'' condition
\be
L_0 \sigma(\delta_{LL}, m_L) \ge 1 .
\label{discoverycond}
\ee
Such a comparison is shown in Fig.~\ref{scaplot} from which it emerges
that: ($i$) 
For the $e^-e^- \to \ell^- e^-$ process 
there is an observable signal in the upper left corner of the  
($\delta_{LL}, m_L$) plane. The extension of this region depends
on $L_0$. 
($ii$) The bound from $\tau \to e \gamma$ does not 
constrain the region of the ($\delta_{LL}, m_L$) plane
compatible with an observable LFV signal and therefore
the reaction $e^-e^- \to \tau^- e^-$ could produce a detectable signal
whithin the highlighted regions of the parameter 
space (upper-left regions in the ($\delta_{LL}, m_L$) plane).
($iii$) As regards the constraints from the $\mu \to e \gamma$ decay
the allowed region in the ($\delta_{LL}, m_L$) plane is shown by 
the circular dark dots (red with colour): 
the process $e^-e^- \to \mu^- e^-$ is 
observable only in a small section of the parameter space
since the allowed region from the $\mu \to e \gamma$ decay 
almost does not overlap with the collider ``discovery'' region except 
for a very small fraction in the case of gaugino masses ($M_1=80$ GeV and 
$M_2=160 $ GeV). The compatibility of values of 
$\delta_{LL} \approx 1$ is due to a cancellation among the  
diagrams that describe the $\ell \to \ell' \gamma$ decay in particular points 
of the parameter space~\cite{illa}.
 
As regards the radiative mechanism that generates the off-diagonal 
elements in mSUGRA models (as discussed in Sec.~\ref{sec2})
one should check if this mechanism may generate large values of 
$\Delta m^2$. The answer is yes, at least for some particular scenario 
of neutrino masses and mixing. It is well known that any `bottom-up' approach that 
reconstructs the $Y_{\nu}$ from the see-saw mechanism and 
neutrino masses and mixings is ambiguous up to a complex, orthogonal 
matrix $R$ ~\cite{casas}. Usually this matrix is taken to be real 
or identical to the unit matrix. However in Ref.~\cite{pascoli} it is shown that in 
the case of a quasi-degenerate neutrino mass spectrum $R$ being complex allows
for values of $\Delta m^2$ being larger by 5-8 orders of magnitude relative 
to the case of $R$ being real or the unit matrix. 
In this case one has~\cite{pascoli}
\be
|(Y^{\dagger}_{\nu}Y_{\nu})_{13}|^{2} \simeq \frac{M^{2}_{R}m^{2}_{\nu}}{v^{4}}
\times {\cal O}(0.1-1.0).
\label{yy}
\ee
Choosing for example 
$M_R=2\cdot 10^{14}$ GeV, $M_{GUT}=2\cdot 10^{16}$ GeV, 
$m_{\nu}=0.3$ eV, $v=174$ GeV in Eq.~(\ref{yy}) 
and $a_{0}=0$, $m_{0}=150$ GeV in Eq.~(\ref{dl2})  
$\Delta m^{2}$ varies in the range
$2400-7800$ GeV$^{2}$, {\it i.e.} 
with $ 100\ \hbox{GeV} \le m_L \le 200\ \hbox{GeV}$,  
$\delta_{LL}$ is in the range ($0.06 - 0.78$).

\section{Summery and Conclusions}
\label{sec5}

The search at lepton colliders of lepton flavour and lepton number violating
signal is complementary to the search for rare leptons decays.
The next generation of linear
colliders will offer the opportunity to look for reactions 
like $e^{\pm}e^{-}\to\ell^{\pm}e^-\; (\ell=\mu,\tau)$
at energies well above the $Z$ peak resonance.
Upper bounds on the cross sections for these processes at the highest
energies reached by LEP, $189\; \mbox{GeV}\leq \sqrt{s} \leq 209\;\mbox{GeV}$,
where given by the OPAL collaboration, Eq.~(\ref{sigmabounds}).

In this paper the reactions
$e^{\pm}e^{-}\to\ell^{\pm}e^-\; (\ell=\mu,\tau)$ induced by 
sleptons mixing in R-parity conserving supersymmetry have been studied.
The reactions proceed through loop diagrams (box and penguin type) 
involving sleptons, neutralinos and charginos. 
The amplitudes have been evaluated in the helicity basis and the loop 
integrals are calculated numerically. 
The resulting cross sections exhibit the well known threshold 
enhancement for center of mass energies corresponding 
to pair production of supersymmetric particles.
In particular due to the dominance of the $(s,t)$-channel box diagrams with 
sleptons on the threshold in the intermediate state, the LFV cross section 
reaches its maximum value at the energy corresponding to the threshold for 
sleptons pair production both in $e^+e^-$ and $e^-e^-$ collisions.    

The $e^-e^-$ option with left-polarized beams stands better 
chances to provide a detectable signal. A comparison with present experimental bounds 
on radiative lepton decays shows that an observable 
($ e^-e^- \to \tau^- e^-$) signal is
compatible with
the non observation of the decay $\tau \to e \gamma$ giving 
some tens of events with an integrated luminosity of 100 fb$^{-1}$. 
On the contrary  
the more restrictive constraints from the non-observation
of $\mu \to e \gamma$ make the search of $ e^-e^- \to \mu^- e^-$
unrealistic unless the integrated luminosity is very large.
It has been shown that the Standard Model background is low and can be easily
suppressed using that the signal final state consists of two back to back high energy 
leptons of different flavour {\em with no missing energy}. 
The observation $e^+e^-$ in collisions will be more difficult because 
of smaller cross sections.    

\begin{acknowledgments}
The work of St. Kolb is supported by the European Union, under contract 
No. HPMF-CT-2000-00752. 

\end{acknowledgments}

\appendix
\section{Lagrangian and couplings}
\label{lagrangian}

The interaction lagrangians in the gauge basis for superparticles
in the notation of Ref.~\cite{Haber}:

\noindent (a){\em Lepton-chargino-sneutrino:} 
\be
{\cal L}=
O^{\tilde{W}}_{\tilde{\nu}}\ovl{\ell}P_R\tilde{W}\tilde{\nu}+h.c.,
\ee
with coupling $O^{\tilde{W}}_{\tilde{\nu}}=-g$.

\noindent (b) {\em Lepton-neutralino-slepton:}
\be
{\cal L}&=&(
O^{\tilde{W}^3}_{\tilde{L}} \ovl{\ell}P_R\tilde{W}^3\tilde{L}
+O^{\tilde{B}}_{\tilde{L}}\ovl{\ell}P_R\tilde{B}\tilde{L}
+O^{\tilde{B}}_{\tilde{R}}\ovl{\ell}P_L\tilde{B}\tilde{R})\cr
&+&h.c.,
\ee
with couplings given by $O^{\tilde{W}^3}_{\tilde{L}}=\frac{g}{\sqrt 2}$,
$ O^{\tilde{B}}_{\tilde{L}}=\frac{g}{\sqrt 2} t_W$, 
$O^{\tilde{B}}_{\tilde{R}}={\sqrt 2}g t_W$ and 
$\tilde{L}\equiv \tilde{\ell}_{L}$, $\tilde{R}\equiv \tilde{\ell}_{R}$.

\noindent (c) {\em Lepton-lepton-vector boson:}
\be
{\cal L}= \sum_{V=\gamma,Z^0}
 V_{\mu}\ovl{\ell}\gamma^{\mu}(O_V^L P_L+O_V^R P_R)\ell
\ee
where $O_{Z^0}^R=-g s_W t_W$, $O_{Z^0}^L=+\frac{g}{c_W}(\frac{1}{2}-s^2_W)$,
$O^{L}_{\gamma}=O^{R}_{\gamma}=e$.

\noindent (d) {\em Slepton-slepton-vector boson:}
\be
{\cal L}=i
O_V^{\tilde{L}\tilde{L}} V_{\mu}\tilde{L}^* \olrap^{\mu} \tilde{L},
\ee
 with $O^{\tilde{\ell}\tilde{\ell}}_{\gamma}=e$, $O^{\tilde{\ell}\tilde{\ell}}_{Z^0} 
=g\frac{g}{c_W}(\frac{1}{2}-s^2_W)$, $O^{\tilde{\nu}\tilde{\nu}}_{Z^0}=-\frac{g}{2c_W}$.

\noindent (e) {\em Chargino-chargino-vector boson:}
\be
{\cal L}=
O_V^{\tilde{W}} V_{\mu} \ovl{\tilde{W}} \gamma^{\mu} \tilde{W}
\ee
with $O_{\gamma}^{\tilde{W}}=-e$, $O_{Z^0}^{\tilde{W}}=-g c_W$.

\section{Numerical tools}
\label{appb} 
\subsection{Spinor products}
\label{appb1}
Here the basic formulas used in the computation of helicity 
amplitudes are given. More details and proofs are given in Ref.~\cite{kleiss}.
The spinor products satisfy exchange relations:
\begin{equation}
\begin{array}{ccc}
S(p_a,p_b)=-S(p_b,p_a), & T(p_a,p_b)=-T(p_b,p_a),\\
S(p_a,p_b)=T^{*}(p_b,p_a), & T(p_a,p_b)=S^{*}(p_a,p_b),\\
|S(p_b,p_a)|^{2}=2 p_a \cdot p_b, & |T(p_a,p_b)|^{2}=2 p_a \cdot p_b.
\label{spinexc}
\end{array}
\end{equation}
The necessary relations to write the amplitudes in terms of spinor products
are the Chisholm identities:
\be
&\left[\ovl{u}_{\lambda}(p_a) {\gamma^{\mu}} u_{\lambda}(p_b)\right]\gamma_{\mu}=&\cr
&2\left[u_{\lambda}(p_b)\ovl{u}_{\lambda}(p_a)
+ u_{-\lambda}(p_a)\ovl{u}_{-\lambda}(p_b)\right],&\cr
&\pslash = u_R(p) \ovl{u}_R(p)+u_L(p) \ovl{u}_L(p).&
\label{spinid}
\ee
where $\lambda=L,R$ indicates the helicity of the spinor.
The external momenta are parametrized in terms of the Mandelstam variable $s$
and the scattering angle in the center of mass frame:
\be
p_1&=&\frac{\sqrt s}{2}(1,0,0,1),\cr
p_2&=&\frac{\sqrt s}{2}(1,0,0,-1),\cr
p_3&=&\frac{\sqrt s}{2}(1,-\sin{\theta},0,-\cos{\theta}),\cr
p_4&=&\frac{\sqrt s}{2}(1,\sin{\theta},0,\cos{\theta}).
\label{momenta}
\ee
The spinor products are determined by the components of these four momenta
in the following way:
\be
S(p_a,p_b)&=&(p_a^z+ip_a^x)\sqrt{\frac{p_b^0-p_b^y}{p_a^0-p_a^y}}\cr
&-&(p_b^z+ip_b^x)\sqrt{\frac{p_a^0-p_a^y}{p_b^0-p_b^y}},
\label{sexplicit}
\ee
and $T(p_a,p_b)$ is easily deduced by ralations~(\ref{spinexc}).
Using Eq.~(\ref{momenta}) and Eq.~(\ref{sexplicit}) it is easy to see that 
the relations~(\ref{spinexc}) are satisfied. In the case of $2 \to 2$ 
scattering, with the momenta given in Eq.~(\ref{momenta}), the preceding 
expressions simplifies to:
\be   
S(p_a,p_b)&=&(p_{a}^{z}-p_{b}^{z})+i(p_{a}^{x}-p_{b}^{x}),\cr
T(p_a,p_b)&=&(p_{b}^{z}-p_{a}^{z})-i(p_{b}^{x}-p_{a}^{x}),
\ee
and product of spinor products are directly related to $s,t,u$. For example
one has:
\be
S(p_1,p_3)T(p_4,p_2)&=&-\frac{s}{2}(1+\cos{\theta})=u,\cr 
S(p_1,p_4)T(p_3,p_2)&=&-\frac{s}{2}(1-\cos{\theta})=t,\cr
S(p_1,p_2)T(p_4,p_3)&=&s e^{i\theta}. 
\ee 
\subsection{Tensor integral decomposition}
\label{appb2}
\begin{figure}
\begin{center}
\scalebox{0.5}{\includegraphics*[75,235][510,735]{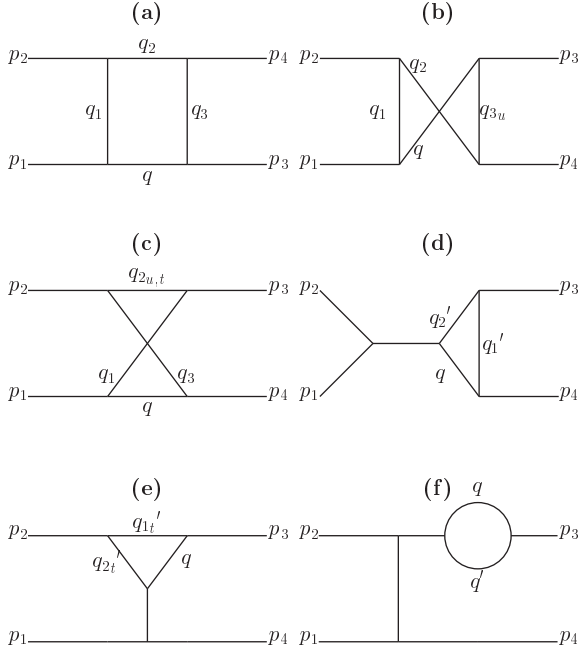}}
\caption{Definition of virtual momenta for kinematics and tensor integral 
decomposition.}
\label{loopkin}
\end{center}       
\end{figure}

The loop integrals are evaluated numerically with the package 
{\scshape{LoopTools}}~\cite{looptools}. Here we report the definitions and
the decomposition for two, three and four point tensor functions.
\be
B_{\mu}&=&\int \frac{d^{4}q}{i\pi^{2}}\frac{q_\mu}{N_1 N_2},\cr
C_{\mu;\alpha\beta}&=&\int \frac{d^{4}q}{i\pi^{2}}\frac{q_\mu; q_{\alpha}
{q_\beta}}{N_1 N_2 N_3},\cr
D_{\mu;\alpha\beta}&=&\int \frac{d^{4}q}{i\pi^{2}}\frac{q_\mu; q_{\alpha}
{q_{\beta}}}{N_1 N_2 N_3 N_4},
\ee
are expressed as:
\be
B_{\mu}&=&k_{1\mu} B_1\cr
C_\mu&=&\sum^{2}_{i=1}{k_{i}}_{\mu}C_{i},\cr   
C_{\mu\nu}&=&g_{\mu\nu}C_{00}+\sum^{2}_{i,j=1}{k_{i}}_{\mu}{k_{j}}_{\nu}C_{ij}\cr
D_\mu&=&\sum^{3}_{i=1}{k_{i}}_{\mu}D_{i},\cr
D_{\mu\nu}&=&g_{\mu\nu}D_{00}+\sum^{3}_{i,j=1}{k_{i}}_{\mu}{k_{j}}_{\nu}D_{ij},
\ee
where the $k_i$'s are sums of external momenta appearing in the loops 
propagators as reported in Fig.~\ref{loopkin}:
\be
q_1&=&q+k_1=q+p_1,\cr
q_2&=&q+k_2=q+p_1+p_2,\cr
{q_2}_{u,t}&=&q+p_1-p_3,\cr
q_3&=&q+k_3=q+p_4,\cr
{q_3}_{u}&=&q+p_3,\cr
q_1'&=&q+k_1'=q+p_4,\cr
q_2'&=&q+k_2'=q+p_3+p_4,\cr
{{q_1}_t}'&=&q+p_3,\cr
{{q_2}_t}'&=&q+p_3-p_2,\cr
q'&=&q-p_3
\label{loopmomenta}
\ee
and the mass and Mandelstam 
variables dependence for a generic two, three, four point function and for 
the various topologies of graphs corresponding to the kinematical channels
are: 
\be
D_{a}&=&D(0,0,0,0,s,t,m^2_{q},m^2_{q_1},m^2_{q_2},m^2_{q_3}),\cr
D_{b}&=&D(0,0,0,0,s,u,m^2_{q},m^2_{q_1},m^2_{q_2},m^2_{q_3}),\cr
D_{c}&=&D(0,0,0,0,u,t,m^2_{q},m^2_{q_1},m^2_{q_2},m^2_{q_3}),\cr
C_{d}&=&C(0,0,s,m^2_{q'},m^2_{q_1'},m^2_{q_2'}),\cr
C_{e}&=&C(0,0,t,m^2_{q'},m^2_{q_1'},m^2_{q_2'}),\cr
B_{f}&=&B(m^2_{q},m^2_{q'}).
\ee

\section{Helicity amplitudes}
\label{amplitudes}
\subsection{$\bm{e^+e^-}$ collisions}
\label{suba}
The amplitudes are given assuming that the negative charged final leptons 
has changed flavour. The other possibility is taken into account simply 
multiplyng the total cross section by two. 
The non-zero helicity amplitudes 
are found to be:

\begin{center}
{\em A: $e^+_R e^-_L \to \ell^-_L e^+_R$}\\ 
\end{center}
For clarity, graphs are grouped according to the virtual particles 
present in the boxes that can be produced in $e^+ e^-$ collisions:\\ 
\noindent {1.\em  Virtual selectrons pair:\\}
There are four box diagrams with all the possible 
$\tilde{B}$ and $\tilde{W}$
assignament in the neutralino lines in Figs.~\ref{boxep}a,~\ref{boxep}b: 
\begin{eqnarray}
{\cal M}_{A,1}^{\Box}&=&
\sum_{i,i'=\tilde{B},\tilde{W^0}} (O^i)^2 (O^{i'})^2 T(p_1,p_3)S(p_4,p_2)\cr
&\times&\left\{ 2D^{ii'}_{00}(s,t)+T(p_1,p_4)S(p_4,p_1)
D_{13}^{i,i'}(s,t)\right. \cr
&-& \left. m_{i}m_{i'} D_{0}^{ii'}(u,t)\right\}. 
\end{eqnarray} 
The terms depending on $(u,t)$ come from the crossed box diagrams due to the 
Majorana nature of neutralinos.
Contribution from $s$ and $t$ channel penguins, Figs.~\ref{penguinep}a,~\ref{penguinep}b
and the corresponding external legs corrections, Fig.~\ref{penguinep}d, give:
\be
{\cal M}_{A,1}^{\triangle}&=& 2 T(p_1,p_3)S(p_4,p_2)\cr
&\times& \left\{\sum_{V,i}D_V(s)\left[
(O^i)^2 O_V^{\tilde{\ell}\tilde{\ell}} O^V_L 2 C_{00}(s)\right.\right.\cr
&-&\left.(O^i)^2 (O^V_L)^{2} (B_{0}+B_{1}) \right] \cr
&-&\left.\sum_{V,i}D_V(t)\left[
(O^i)^2 O_V^{\tilde{\ell}\tilde{\ell}} O^V_L 2 C_{00}(t)\right.\right.\cr
&-&\left.\left.(O^i)^2 (O^V_L)^{2}(B_{0}+B_{1})\right]\right\} 
\ee
The photon and $Z^{0}$ propagators are given by 
$D_V(s)=-i/(s-M^{2}_{V}+iM_{V}\Gamma_{V})$ ($V=\gamma,Z)$ for the $s$-channel,
while no immaginary part is present in the denominator for $t$ and $u$
channels. 

\noindent{\em 2. Virtual sneutrinos pair:\\}
The box diagrams in Fig.~\ref{boxep}c reads:
\be 
{\cal M}_{A,2}^{\Box}&=&
({O^{\tilde{W}}_{\tilde{\nu}}})^4
T(p_1,p_3)S(p_4,p_2)\left\{2 D_{00}(s,t)\right.\cr
&+&\left.T(p_1,p_4)S(p_1,p_4)D_{13}(s,t)\right\},
\ee
while the penguins of Figs.~\ref{penguinep}b,~\ref{penguinep}e:
\be
{\cal M}_{A,2}^{\triangle}&=& 2 T(p_1,p_3)S(p_4,p_2)\cr
&\times&\left\{ D_Z(s)\left[
({O^{\tilde{W}}_{\tilde{\nu}}})^2 O_Z^{\tilde{\nu}\tilde{\nu}} O^Z_L 2 C_{00}(s)\right.\right.\cr
&-&\left.({O^{\tilde{W}}_{\tilde{\nu}}})^2 (O^Z_L)^{2}(B_{0}+B_{1})\right]\cr
&-&\left.D_Z(t)\left[({O^{\tilde{W}}_{\tilde{\nu}}})^2 
O_Z^{\tilde{\nu}\tilde{\nu}} O^Z_L 2 C_{00}(t)\right.\right.\cr
&-&\left.\left.({O^{\tilde{W}}_{\tilde{\nu}}})^2 (O^Z_L)^{2}(B_{0}+B_{1})\right]\right\}
\ee
The amplitudes present the same structure of those in case 1.

\noindent{\em 3. Virtual chargino pair:\\}

The box diagram in Fig.~\ref{boxep}d:

\be 
 {\cal M}_{A,3}^{\Box} 
&=&({O^{\tilde{W}}_{\tilde{\nu}}})^{4}T(p_1,p_3)S(p_2,p_4)\left\{2D_{00}(s,t)\right.\cr
&+&\left.T(p_1,p_2)S(p_1,p_2){\cal D}(s,t)\right\}.
\ee

where $k,l=s,t,u$ and
\begin{equation}
{\cal D}(k,l)=[D_{12}(k,l)+D_{22}(k,l)+D_{23}(k,l)+D_2(k,l)], 
\nonumber
\end{equation}

The penguin diagrams in Figs.~\ref{penguinep}c,~\ref{penguinep}e give:

\be
{\cal M}_{A,3}^{\triangle}&=& 2T(p_1,p_3)S(p_4,p_2)\cr
&\times& \left\{  
\sum_{V=\gamma,Z^0}D_V(s)\left[({O^{\tilde{W}}_{\tilde{\nu}}})^{2}
 O_V^{\tilde{W}} O^V_L {\cal C}(s)\right.\right.\cr  
&-&\left.\left.({O^{\tilde{W}}_{\tilde{\nu}}})^{2}(O^{\gamma}_L)^{2}(B_0+B_1)
\right] \right. \cr
&-&\left.
\sum_{V=\gamma,Z^0}D_V(t)\left[({O^{\tilde{W}}_{\tilde{\nu}}})^{2}
 O_V^{\tilde{W}} O^V_L {\cal C}(t)\right.\right.\cr 
&-&\left.\left.({O^{\tilde{W}}_{\tilde{\nu}}})^{2}(O^{\gamma}_L)^{2}(B_0+B_1)
\right] \right\}
\ee
where $k=s,t,u$ and

\be
{\cal C}(k)&=&\left\{ C_0(k) m^2_{\tilde{W}}\right.\cr
&-&\left.\left[ 2 C_{00}(k)
+k(C_{2}(k)+C_{12}(k)+C_{22}(k))\right]\right\}.
\nonumber
\ee

\noindent{\em 4. Virtual neutralino pair:\\} 
There are four box diagrams with left-slepton and all possible combinations
of Bino and neutral Wino in the loop of Figs.~\ref{boxep}e,~\ref{boxep}f with left 
sleptons exchanged: 

\be 
{\cal M}_{A,4}^{\Box}&=&\sum_{i,i'}(O^i)^2 (O^{i'})^2 T(p_1,p_3)S(p_2,p_4)\cr
&\times& \left\{ 2D^{ii'}_{00}+T(p_1,p_2)S(p_1,p_2){\cal D}^{i,i'}(s,t)\right.\cr
&-&\left.m_{i}m_{i'} D_{0}^{ii'}(s,u)\right\}  
\ee 

Note that there is no 
penguin contribution to this channel in the $\tilde{B},\tilde{W}^0$ 
basis. 
The amplitudes ${\cal M}^{\Box}_{A,3}$ and ${\cal M}^{\Box}_{A,4}$ have a 
minus sign relative to the other amplitudes, because 
$T(p_1,p_3)S(p_2,p_4)=-T(p_1,p_3)S(p_4,p_2)$, see Eq.~(\ref{spinexc}).
Its origin is due to the fact that once one fixes the order of the spinors,
the two different topologies of box diagrams need an odd number of permutation 
of fermion fields to bring them in the same order. The same holds for 
the relative sign between $s$ and $t$ channel penguins.

\begin{center}
{\em B: $e^+_L e^-_R \to \ell^-_L e^+_R$}\\
\end{center}
This differs from the previous by the exchange of initial state helicity:
only the penguin diagrams contribute and the amplitudes are obtained 
selecting the $O^{\gamma,Z}_R P_R$ operator in the lepton-lepton-vector boson 
vertex. 

\be
{\cal M}_{B,1}^{\triangle}& =& 2 T(p_1,p_4)S(p_3,p_2)\cr
&\times& \left\{ \sum_{V,i}D_V(s)\left[
(O^i)^2 O_V^{\tilde{\ell}\tilde{\ell}} O^V_R 2 C_{00}(s)\right.\right.\cr
&-&\left.\left.(O^i)^2 (O^V_R)^{2}(B_{0}+B_{1})\right]\right. \cr
&-&\left.\sum_{V,i}D_V(t)\left[(O^i)^2 O_V^{\tilde{\ell}\tilde{\ell}} O^V_R 2 C_{00}(t)\right.\right.\cr
&-&\left.\left.(O^i)^2 (O^V_R)^{2}(B_{0}+B_{1})\right] \right\} 
\ee

\be
{\cal M}_{B,2}^{\triangle}&=&2 T(p_1,p_4)S(p_3,p_2)\cr
&\times& \left\{D_Z(s)\left[ ({O^{\tilde{W}}_{\tilde{\nu}}})^{2}
 O_Z^{\tilde{\nu}\tilde{\nu}} O^Z_R 2 C_{00}(s)\right.\right.\cr
&-&\left.\left.({O^{\tilde{W}}_{\tilde{\nu}}})^{2}(O^Z_R)^{2}(B_{0}+B_{1})\right] \right. \cr
&-& \left. D_Z(t)\left[ 
({O^{\tilde{W}}_{\tilde{\nu}}})^{2}
 O_Z^{\tilde{\nu}\tilde{\nu}} O^Z_R 2 C_{00}(t)\right.\right.\cr
&-&\left.\left.({O^{\tilde{W}}_{\tilde{\nu}}})^{2}(O^Z_R)^{2}(B_{0}+B_{1})\right] \right\}
\ee

\be
{\cal M}_{B,3}^{\triangle}&=& 
2T(p_1,p_4)S(p_3,p_2)\cr
&\times& \left\{
\sum_{V=\gamma,Z^0}D_V(s)\left[({O^{\tilde{W}}_{\tilde{\nu}}})^{2}
O_V^{\tilde{W}} O^V_R {\cal C}(s)\right.\right.\cr
&-&\left.\left.({O^{\tilde{W}}_{\tilde{\nu}}})^{2}(O^{\gamma}_R)^{2}(B_0+B_1)\right] \right.\cr
&-&\left.
\sum_{V=\gamma,Z^0}D_V(t)\left[({O^{\tilde{W}}_{\tilde{\nu}}})^{2}
 O_V^{\tilde{W}} O^V_R {\cal C}(t)\right.\right.\cr
&-&\left.\left.({O^{\tilde{W}}_{\tilde{\nu}}})^{2}(O^{\gamma}_R)^{2}(B_0+B_1)
\right] \right\}
\ee

\begin{center}
{\em C: $e^+_L e^-_L \to \ell^-_L e^+_L$} \\
\end{center}

The box diagram in Figs.~\ref{boxep}e,~\ref{boxep}f with the right-handed 
selectron and the Binos in the neutralinos lines contributes:

\be
{\cal M}_{C,1}^{\Box}&=& 
({O^{\tilde{B}}_{\tilde{L}}})^{2}({O^{\tilde{B}}_{\tilde{R}}})^{2}
\left\{ S(p_1,p_2)T(p_3,p_4)2D_{00}(s,t)\right. \cr
&+&\left. S(p_1,p_2)T(p_2,p_4)T(p_3,p_1)S(p_1,p_2){\cal D}(s,t)\right. \cr
&-& \left. S(p_1,p_2)T(p_3,p_4)\left[2D_{00}(s,u)\right.\right.\cr 
&+&\left.\left.S(p_1,p_4)T(p_1,p_4)D_{13}(s,u)\right]\right\}
\ee

The box diagrams in Figs.~\ref{boxep}g,~\ref{boxep}h:

\be
{\cal M}_{C,2}^{\Box}&=&
({O^{\tilde{B}}_{\tilde{L}}})^{2}({O^{\tilde{B}}_{\tilde{R}}})^{2}
S(p_1,p_2)T(p_3,p_4)\cr
&\times& \left\{ m_{\tilde{B}}^2 D_0(s,t)-2D_{00}(u,t)\right. \cr
&+&\left.S(p_1,p_4)T(p_1,p_4)D_{13}(u,t)\right\}
\ee

\subsection{$\bm{e^{-}e^{-}}$ collisions}
\label{subb}

The helicity amplitudes are:
\begin{center}
{\em E1: $e^-_L e^-_L \to \ell^-_L e^-_L$}\\
\end{center}
Four box diagrams of the kind given in Fig.~\ref{figee}a with left
sleptons and the box in Fig.~\ref{figee}b with charginos:

\be
&{\cal M}^{\Box}_{E1}&= 
\sum_{i,i'}
(O^i)^2 (O^{i'})^2 S(p_1,p_2)T(p_4,p_3)\cr
&\times & \left\{ m_{i}m_{i'} D_{0}^{i}(s,t)+2 D_{00}^{ii'}(u,t)\right. \cr
&-&\left.S(p_4,p_1)T(p_4,p_1)D_{13}^{ii'}(u,t)\right\}\cr
&+& ({O^{\tilde{W}}_{\tilde{\nu}}})^4
\left[2D^{c}_{00}(u,t)+S(p_4,p_1)T(p_4,p_1)D^{c}_{13}(u,t)\right] \cr
&-& {\cal M}(p_{1}\leftrightarrow p_{2}, u \leftrightarrow t)
\ee

Penguin diagrams in $t$ and $u$ channels, with left couplings with gauge 
bosons:

\be
{\cal M}_{E1,1}^{\triangle}&=& 2 S(p_2,p_1)T(p_4,p_3)\cr
&\times&\left\{  \sum_{V,i}D_V(t)\left[
(O^i)^2 O_V^{\tilde{\ell}\tilde{\ell}} O^V_L 2 C_{00}(t)\right.\right.\cr
&-&\left.\left.(O^i)^2 (O^V_L)^{2}(B_{0}+B_{1})\right]\right.\cr
&-&\left.\sum_{V,i}D_V(u) \left[(O^i)^2 O_V^{\tilde{\ell}\tilde{\ell}} O^V_L 2 C_{00}(u)\right.\right.\cr
&-&\left.\left.(O^i)^2 (O^V_L)^{2}(B_{0}+B_{1})\right]\right\}\cr
&-& {\cal M}(p_{1}\leftrightarrow p_{2}, u \leftrightarrow t) 
\ee

\be
{\cal M}_{E1,2}^{\triangle}&=&
2 S(p_2,p_1)T(p_4,p_3)\cr
&\times& \left\{ D_V(t) \left[
({O^{\tilde{W}}_{\tilde{\nu}}})^{2} O_V^{\tilde{\nu}\tilde{\nu}} O^Z_L 2 C_{00}(t)\right.\right.\cr
&-&\left.\left.(O^i)^2 (O^Z_L)^{2}(B_{0}+B_{1})\right]\right.\cr
&-&\left.
 D_Z(u) \left[
({O^{\tilde{W}}_{\tilde{\nu}}})^{2}
 O_Z^{\tilde{\nu}\tilde{\nu}} O^Z_L 2 C_{00}(u)\right.\right.\cr
&-&\left.\left.
({O^{\tilde{W}}_{\tilde{\nu}}})^{2}(O^Z_L)^{2}(B_{0}+B_{1})\right]\right\}\cr
&-& {\cal M}(p_{1}\leftrightarrow p_{2}, u \leftrightarrow t)
\ee

\be
{\cal M}_{E1,3}^{\triangle}&=& 
2S(p_2,p_1)T(p_4,p_3)\cr
&\times&\left\{ 
\sum_{V=\gamma,Z^0}D_V(t)\left[({O^{\tilde{W}}_{\tilde{\nu}}})^{2}
 O_V^{\tilde{W}} O^V_L {\cal C}(t)\right.\right.\cr
&-&\left.\left.({O^{\tilde{W}}_{\tilde{\nu}}})^{2}(O^{\gamma}_L)^{2}(B_0+B_1)
\right]\right. \cr
&-&\left.
\sum_{V=\gamma,Z^0}D_V(u)\left[({O^{\tilde{W}}_{\tilde{\nu}}})^{2}
 O_V^{\tilde{W}} O^V_L {\cal C}(u)\right.\right.\cr
&-&\left.\left.({O^{\tilde{W}}_{\tilde{\nu}}})^{2}(O^{\gamma}_L)^{2}(B_0+B_1)
\right] \right\}\cr
&-& {\cal M}(p_{1}\leftrightarrow p_{2}, u\leftrightarrow t)
\ee
All amplitudes are anty-symmetrized respect to initial state
identical leptons.

\begin{center}
{\em E2: $e^-_L e^-_R \to \ell^-_L e^-_R$} \\
\end{center}
Box diagrams of Fig.~\ref{figee}e and penguins with left coupling 
of gauge bosons to leptons:

\be
{\cal M}^{\Box}_{E2}&=& 
({O^{\tilde{B}}_{\tilde{L}}})^{2}({O^{\tilde{B}}_{\tilde{R}}})^{2}
T(p_1,p_3)S(p_4,p_2)\cr
&\times& \left[2 D_{00}(s,t)+T(p_1,p_4)S(p_4,p_2)D_{31}(s,t)\right.\cr
&-&\left.m_{\tilde{B}}^{2}D_{0}(u,t)\right]
\ee

\be
{\cal M}_{E2,1}^{\triangle}&=& 
2 T(p_1,p_3)S(p_4,p_2)\cr
&\times& \sum_{V,i}D_V(t)\left\{
(O^i)^2 O_V^{\tilde{\ell}\tilde{\ell}} O^V_R 2 C_{00}(t)\right.\cr
&-&\left.(O^i)^2 (O^V_R)^{2}(B_{0}+B_{1})\right\} 
\ee

\be
{\cal M}_{E2,2}^{\triangle}&=&
2 T(p_1,p_3)S(p_4,p_2)\cr
&\times& D_Z(t)\left\{ 
({O^{\tilde{W}}_{\tilde{\nu}}})^{2} 
O_Z^{\tilde{\nu}\tilde{\nu}} O^Z_R 2 C_{00}(t)\right.\cr
&-&\left.({O^{\tilde{W}}_{\tilde{\nu}}})^{2}(O^Z_R)^{2}(B_{0}+B_{1})\right\}
\ee

\be
{\cal M}_{E2,3}^{\triangle}&=&
2T(p_1,p_3)S(p_4,p_2)\cr
&\times& 
\sum_{V=\gamma,Z^0}D_V(t)\left\{({O^{\tilde{W}}_{\tilde{\nu}}})^{2}
 O_V^{\tilde{W}} O^V_R {\cal C}(t)\right.\cr
&-&\left.({O^{\tilde{W}}_{\tilde{\nu}}})^{2}(O^{\gamma}_R)^{2}(B_0+B_1)\right\}
\ee

\begin{center}
{\em E3: $e^-_R e^-_L \to \ell^-_L e^-_R$} \\
\end{center}
This is obtained simply exchanging $p_1\leftrightarrow p_2$ 
and $t \leftrightarrow u$ in the previous amplitudes.

\be
{\cal M}^{\Box}_{E3}&=&
({O^{\tilde{B}}_{\tilde{L}}})^{2}({O^{\tilde{B}}_{\tilde{R}}})^{2}
T(p_2,p_3)S(p_4,p_1)\cr
&\times& \left[2D_{00}(s,u)+T(p_2,p_4)S(p_4,p_1)D_{31}(s,u)\right.\cr
&-&\left.m_{\tilde{B}}^{2}D_{0}(t,u)\right]
\ee

\be
{\cal M}_{E3,1}^{\triangle}&=& 
2 T(p_2,p_3)S(p_4,p_1)\cr
&\times& \sum_{V,i}D_V(u)\left\{
(O^i)^2 O_V^{\tilde{\ell}\tilde{\ell}} O^V_R 2 C_{00}(u)\right.\cr
&-&\left.(O^i)^2 (O^V_R)^{2}(B_{0}+B_{1})\right\} 
\ee

\be
{\cal M}_{E3,2}^{\triangle}&=&
2 T(p_2,p_3)S(p_4,p_1)\cr
&\times& D_Z(u)\left\{ 
({O^{\tilde{W}}_{\tilde{\nu}}})^{2}
 O_Z^{\tilde{\nu}\tilde{\nu}} O^Z_R 2 C_{00}(u)\right.\cr
&-&\left.({O^{\tilde{W}}_{\tilde{\nu}}})^{2}(O^Z_R)^{2}(B_{0}+B_{1})\right\}
\ee

\be
{\cal M}_{E3,3}^{\triangle}&=&
2T(p_2,p_3)S(p_4,p_1)\cr
&\times& 
\sum_{V=\gamma,Z^0}D_V(u)\left\{({O^{\tilde{W}}_{\tilde{\nu}}})^{2}
 O_V^{\tilde{W}} O^V_R {\cal C}(u)\right.\cr
&-&\left.({O^{\tilde{W}}_{\tilde{\nu}}})^{2}(O^{\gamma}_R)^{2}(B_0+B_1)\right\}.
\ee

Some important remarks: each diagram with a LFV and a LFC scalar line is described
by the propagators of Eq.~(\ref{LFVprop}) and Eq.~(\ref{LFCprop}), so that the loop coefficients
in the amplitudes are a sum of four integrals, while the ones with only
the LFV line is a sum of two.  
The scalar two point function $B_0$ and the tensor coefficients 
$B_1$, $C_{00}$ that appear in 
the electroweak penguins are ultra-violet divergent, but the amplitudes are 
finite due the ortogonality of the slepton mixing matrix. 

Penguin diagrams with the exchange of the photon in the $t$ or $u$ channel
are divergent for $t,u \to 0$. This divergence is cancelled by the 
graphs with external legs renormalization as required by gauge invariance.   
As explained in Fig.~\ref{penguinep}, the $t$-channel penguin diagrams
where a scalar line is not dotted, contribute two times because the LFV
propagator may appear once in both lines. 
The two amplitudes are equal because of 
the symmetry property of {\scshape{LoopTools}} form factors giving in this way 
a factor of two, that is necessary for the cancellation of the small $t$ or $u$
divergence.   
Finally, each amplitude gets a factor 
$\frac{i\pi^{2}}{(2\pi)^4}=i\frac{1}{(4\pi)^2}$
from the loop normalization convention.


\begin{thebibliography}{99}


\bibitem{Gabbiani}
F.~Gabbiani, E.~Gabrielli, A.~Masiero and L.~Silvestrini,
Nucl.\ Phys.\ B {\bf 477}, 321 (1996)


\bibitem{mega}
M.~Ahmed {\it et al.} [MEGA collaboration],
Phys.\ Rev.\ D {\bf 65}, (2002) 112002

\bibitem{cleo1}
K.~W.~Edwards {\it et al.}  [CLEO Collaboration],
Phys.\ Rev.\ D {\bf 55}, (1997) 3919.

\bibitem{cleo2}
S.~Ahmed {\it et al.}  [CLEO Collaboration],
Phys.\ Rev.\ D {\bf 61}, (2000) 071101

\bibitem{pdg}
D.~E.~Groom {\it et al.}  [Particle Data Group Collaboration],
Eur.\ Phys.\ J.\ C {\bf 15}, (2000) 1.

\bibitem{desy}
J.~A.~Aguilar-Saavedra {\it et al.}  [ECFA/DESY LC Physics Working Group
                  Collaboration], hep-ph/0106315.

\bibitem{illa}
J.~I.~Illana and M.~Masip,
Phys.\ Rev.\ D {\bf 67} (2003) 035004

\bibitem{opal}
G.~Abbiendi {\it et al.}  [OPAL Collaboration],
Phys.\ Lett.\ B {\bf 519}, (2001) 23


\bibitem{lfvee}
N.~V.~Krasnikov, Phys.\ Lett.\ B {\bf 388}, (1996) 783;\\
N.~Arkani-Hamed, H.~C.~Cheng, J.~L.~Feng and L.~J.~Hall,
Phys.\ Rev.\ Lett.\  {\bf 77}, (1996) 1937; 
Nucl.\ Phys.\ B {\bf 505}, (1997) 3;\\
M.~Hirouchi and M.~Tanaka,
Phys.\ Rev.\ D {\bf 58}, (1998) 032004;\\
J.~Hisano, M.~M.~Nojiri, Y.~Shimizu and M.~Tanaka,
Phys.\ Rev.\ D {\bf 60}, (1999) 055008;\\
D.~Nomura,
Phys.\ Rev.\ D {\bf 64}, (2001) 075001;\\
M.~Guchait, J.~Kalinowski and P.~Roy,
Eur.\ Phys.\ J.\ C {\bf 21}, (2001) 163;\\
W.~Porod and W.~Majerotto,
Phys.\ Rev.\ D {\bf 66}, (2002) 015003


\bibitem{Cannoni}
M.~Cannoni, S.~Kolb and O.~Panella,
Eur.\ Phys.\ J.\ C {\bf 28}, 375 (2003)


\bibitem{borma}
F.~Borzumati and A.~Masiero, Phys.\ Rev.\ Lett.\  {\bf 57}, (1986) 961; 

\bibitem{hisa1}
J.~Hisano, T.~Moroi, K.~Tobe, M.~Yamaguchi, T.~Yanagida,
Phys.\ Lett.\ B {\bf 357}, (1995) 579;

\bibitem{hisa2}
J.~Hisano, T.~Moroi, K.~Tobe, M.~Yamaguchi,
Phys.\ Rev.\ D {\bf 53}, (1996) 2442;

\bibitem{casas}
J.~A.~Casas and A.~Ibarra,
Nucl.\ Phys.\ B {\bf 618}, (2001) 171

\bibitem{hisa3}
J.~Hisano and D.~Nomura,
Phys.\ Rev.\ D {\bf 59}, (1999) 116005

\bibitem{biqi}
X.~J.~Bi, Y.~B.~Dai and X.~Y.~Qi,
Phys.\ Rev.\ D {\bf 63}, (2001) 096008

\bibitem{masiso10}
A.~Masiero, S.~K.~Vempati and O.~Vives,
Nucl.\ Phys.\ B {\bf 649}, 189 (2003)

\bibitem{ellis}
J.~R.~Ellis, M.~E.~Gomez, G.~K.~Leontaris, S.~Lola and D.~V.~Nanopoulos,
Eur.\ Phys.\ J.\ C {\bf 14}, (2000) 319

\bibitem{pas}
F.~Deppisch, H.~Pas, A.~Redelbach, R.~Ruckl and Y.~Shimizu,
Eur.\ Phys.\ J.\ C {\bf 28}, 365 (2003)

\bibitem{looptools} 
T.~Hahn and M.~Perez-Victoria, 
Comput.\ Phys.\ Commun.\ {\bf 118}, 153 (1999);\\
http://www.feynarts.de/looptools. 

\bibitem{Peskin}
M.~E.~Peskin,
Int.\ J.\ Mod.\ Phys.\ A {\bf 13}, 2299 (1998).

\bibitem{Heusch}
C.~A.~Heusch and P.~Minkowski,
Nucl.\ Phys.\ B {\bf 416}, 3 (1994).

\bibitem{comphep}
A.~Pukhov {\it et al.}, hep-ph/9908288.

\bibitem{Altarelli}
G.~Altarelli, F.~Caravaglios, G.~F.~Giudice, P.~Gambino and G.~Ridolfi,
JHEP {\bf 0106}, 018 (2001)

\bibitem{pascoli}
S.~Pascoli, S.~T.~Petcov and C.~E.~Yaguna, hep-ph/0301095.

\bibitem{Haber}
H.~E.~Haber and G.~L.~Kane,
Phys.\ Rept.\  {\bf 117}, 75 (1985).

\bibitem{kleiss}
R.~Kleiss and W.~J.~Stirling,
Nucl.\ Phys.\ B {\bf 262} (1985) 235.



\end{thebibliography}
\end{document}